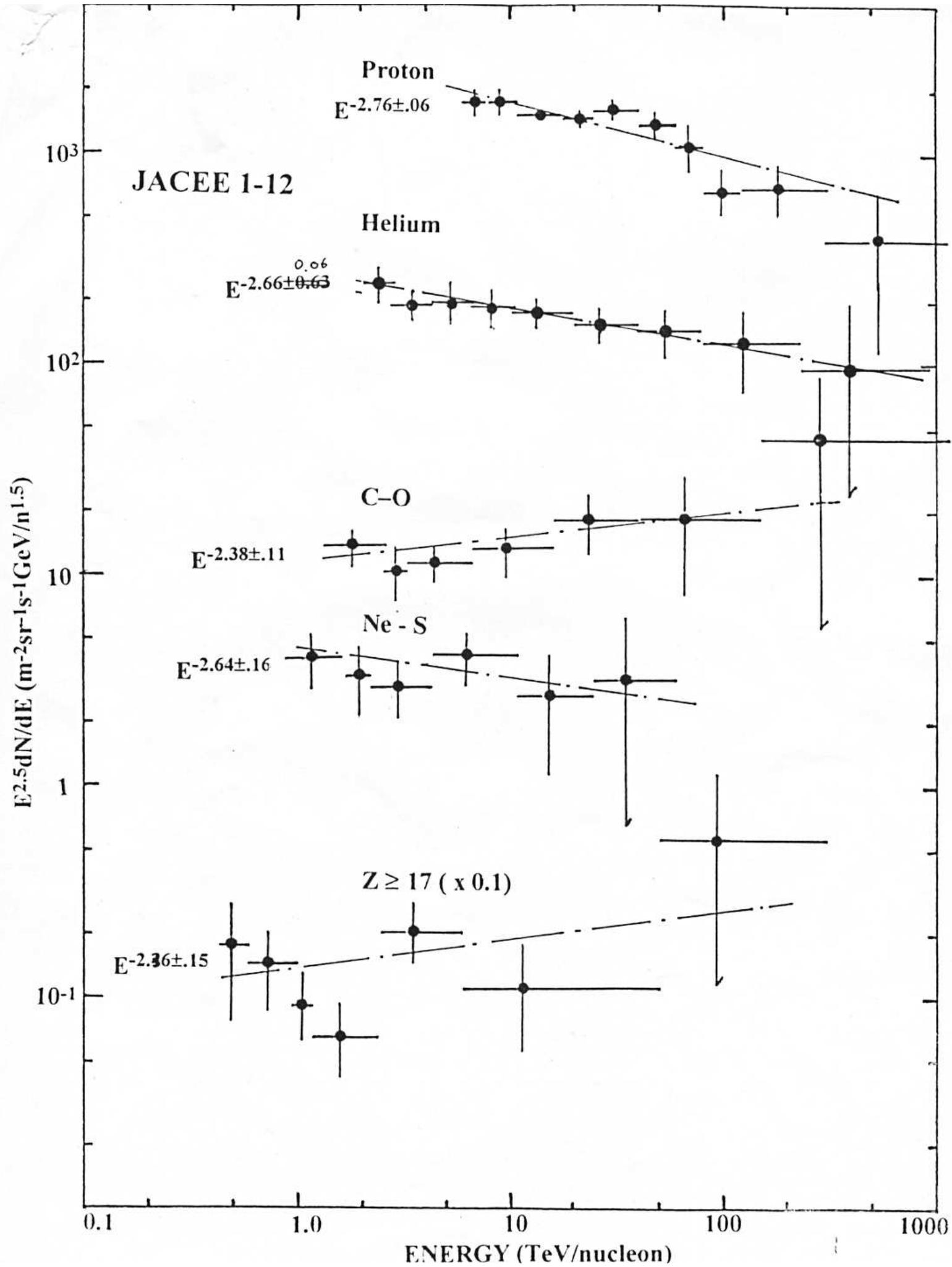

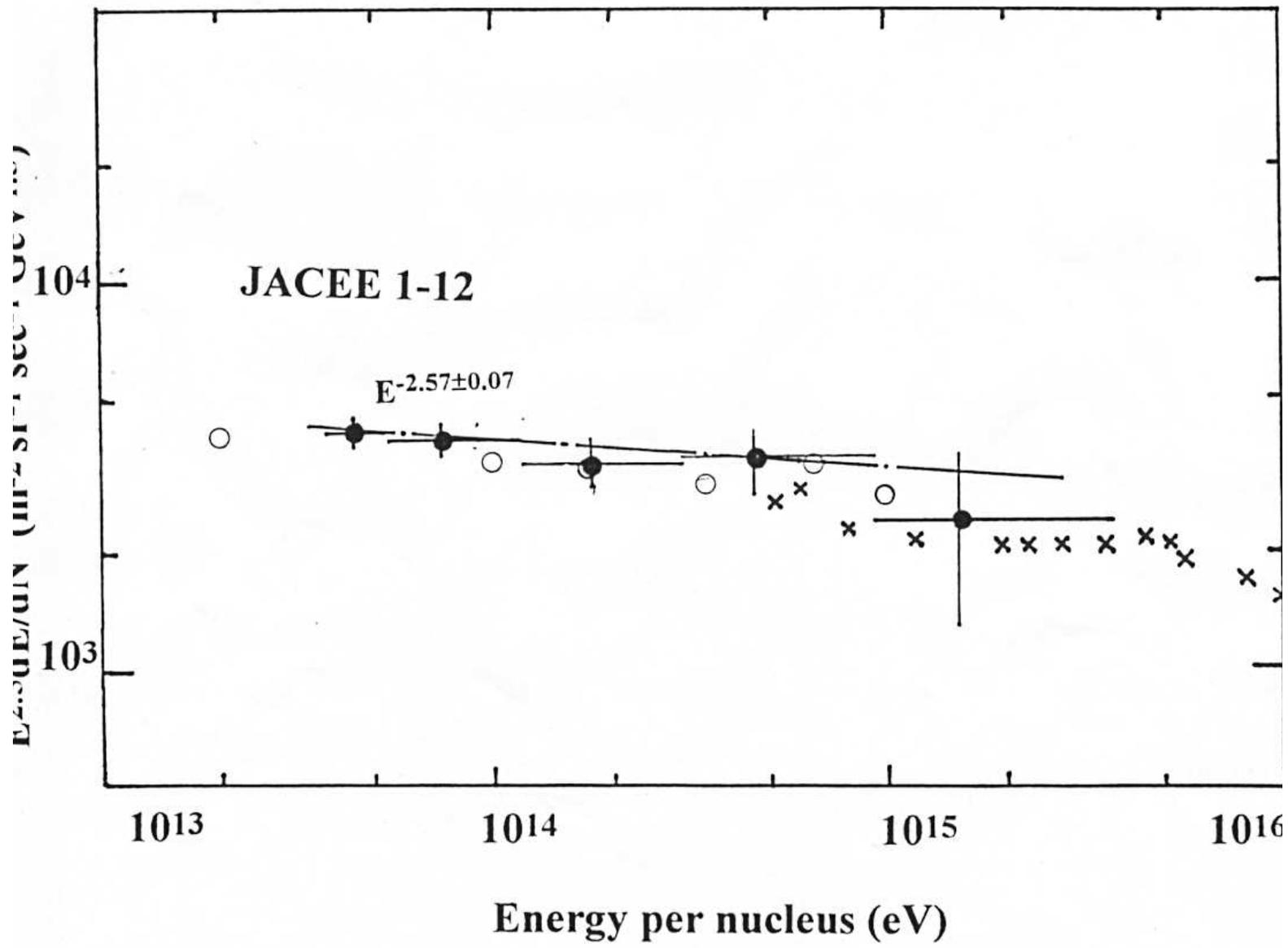

The " all-particle" cosmic ray spectrum in total energy

# Energy Spectra and Elemental Composition of Nuclei above 100 TeV from a Series of the JACEE Balloon Flight.

## The JACEE Collaboration

K.Asakimori[i], T.H.Burnett[d], M.L.Cherry[b], K.Chevli[a], M.J.Christl[c], S.Dake[h],
J.H.Derrickson[c], W.F.Fountain[c], M.Fuki[j], J.C.Gregory[a], T.Hayashi[n], R.Holynski[e],
J.Iwai[d], A.Iyono[l], J.Johnson[a], W.V.Jones[b], M.Kobayashi[k], J.J.Lord[d], O.Miyamura[g],
K.H.Moon[c], H.Oda[h], T.Ogata[f], E.D.Olson[d], T.A.Parnell[c], F.E.Roberts[c], K.Sengupta[b],
T.Shiina[a], S.C.Strausz[d], T.Sugitate[g], Y.Takahashi[a], T.Tominaga[a], J.W.Watts[c],
J.P.Wefel[b], B.Wilczynska[e], H.Wilczynski[e], R.J.Wilkes[d], W.Wolter[e], H.Yokomi[m], and
E.L.Zager[d].

[a]University of Alabama in Huntsville, [b]Louisiana State University, [c]NASA/Marshall Space Flight Center, [d]University of Washington, [e]Institute of Nuclear Physics, Krakow, [f]Institute for Cosmic Ray Research, University of Tokyo, [g]Hiroshima University, [h]Kobe University, [i]Kobe Women's Junior College, [j]Kochi University, [k]National Laboratry for High Energy Physics(KEK), [l]Okayama University of Science, [m]Tezukayama University, [n]Waseda University.

## Abstract

The Japanese American Cooperative Emulsion Experiments (JACEE) have recently carried out a series of Antarctic circumpolar long duration balloon flights (JACEE-10 ~ JACEE-13) for the study of high energy elemental composition and energy spectra of cosmic ray particles. The total exposure factor of these experiments is 664 m$^2$hr, which corresponds to about twice the cumulative exposure from JACEE-1 to JACEE-8. The preliminary result of JACEE-12 analysis shows 8, 5, 4 events for charge groups, C–O, Ne– S, and Z≥17, respectively, above the selection energy of $\Sigma E\gamma \approx$ 20 TeV.

## 1. Introduction

The elemental composition and the energy spectra of cosmic rays around the "knee" region ($10^{15}$ ~$10^{16}$ eV/nucleus) have been central issues in the study of the acceleration mechanism and the origin of cosmic rays. First order Fermi acceleration in shock waves in supernova remnants has been considered to be a plausible acceleration mechanism up to particle energy of $10^{14-15}$ eV. However, for >$10^{14-15}$eV, Fermi acceleration becomes less effective and some other mechanisms or even "new sources" might be necessary to account for the observed slope change and the intensity enhancement in the total particle spectrum around the "knee" [(1)(2)].

Since 1979, JACEE has conducted 14 balloon flights, including 5 long duration flights. We have previously summarized the results of the composition measurements from the first nine flights that include two long duration flights (5-7 days) from Australia to South America [(3)]. In short, the average mass of the cosmic rays in the data set increased with increasing energy toward the "knee", due in part to a deficiency of proton in the observed intensity, and the observed harder energy spectra of other heavy nuclei above 50TeV/nucleus. Additional data analysis above 50 TeV/nucleus from the latest long-duration balloon experiments (JACEE-10 through -13) is currently in progress. The present paper describes the analysis techniques in the presence of the high background densities encountered in long duration Antarctic flights.

## 2. Experimental Techniques and Results

The latest long duration experiments are Antarctic flights which permitted exposures much longer than those of the previous balloon flights. JACEE made four Antarctic flights (JACEE-10~13) and recovered three of them. (JACEE-11 is not recovered yet and still under the Antarctic Sea ). The detail of these flights are reported in another paper in this conference [4]. Total exposure factor of these three experiments is 664 m$^2$hr, corresponding to about double the total exposure factor of JACEE 1-8. We have recently completed the analysis of data from JACEE-10, and the results are presented in [5].

In Antarctic long duration balloon experiments, the low cut-off rigidity produces a radiation background that is expected to cause contamination of emulsion chamber measurements.

When JACEE-10 flew in 1990, it was near "solar maximum" period, and the background of low energy galactic cosmic rays were at a minimum, yielding no effective contamination in the analysis. However, JACEE-12 and -13 were flown in 1994 during near "solar minimum" conditions in which the high background density of low energy particles affected the regular analysis of X-ray films and emulsion plates. These analyses include tracing mesons (particles produced in interactions) upward in the emulsion layers until the interaction vertex is found, measuring the primary charge by "grain" and "$\delta$-ray" counting in the emulsion, and measuring the optical density of X-ray film dark spots to determine the energy.

The tracing of events in the emulsion plates was carried out by the triangulation method using several background heavy tracks, which can locate the position of the events in each emulsion plate within ~10 micron accuracy. This method allowed us to identify uniquely even a single charged track under the high background conditions.

In the charge measurements of Light to Medium Nuclei (Z=3~8), the grain density counting method in the low sensitivity and thick emulsion plate (Fuji ET6B, 150 $\mu$m) was utilized to reduce possible contamination of background tracks in $\delta$-ray counting.

For Heavy Nuclei (Z$\geq$10), $\delta$-ray range distribution measurements were performed, instead of the conventional procedure of $\delta$-ray counting [7], which was limited to resolution $\sigma_z$=1~2 because of uncertainties of the criterion on $\delta$-ray range. The background contamination is neglible in $\delta$-ray range measurements for Heavy Nuclei , so the high sensitivity emulsion plates (Fuji ET7B) have been used. The charge resolution by $\delta$-ray range measurement is evaluated at $\sigma_z \approx 0.3$ for Z$\geq$ 14 by the measurement of background nuclei, which were detected by general scanning in JACEE-10 emulsion plates [8].

The background optical density in the X-ray films ($D_{bg}$) of JACEE-12 was ~1.6, which was higher than $D_{bg}$ of our previous balloon flights by 0.5 ~ 1.0. Since high $D_{bg}$ limits the effective dynamic range of the optical density of the events ($D_{net} = D_{observe} - D_{bg}$), we used a filter to shift the dynamic range higher by 1.4 in the measurements of high energy events. A further discussion of the energy determination is given in [6].

Presently, the tracing of about 200 high energy events from JACEE-12 (5 of total 6 chambers) has been completed and the primary charge of each event has been determined. Preliminary results from JACEE-12 analysis indicate that the number of events of each charge group, C–O, Ne–S, and $Z \geq 17$ are 8, 5, 4 events, respectively, above the selection threshold energy $\Sigma E_\gamma \approx 20$ TeV. Table 1 shows the number of events, and compares the statistics to previous exposures. The proton and helium results are reported in an accompanying paper [6].

JACEE-13, flown in December, 1994, is also being analyzed. Further results of the analysis, including the data from JACEE-12 and the highest energy data from JACEE-13, will soon become available.

## 3. Acknowledgments

This work has been supported in Japan by the Institute for Cosmic Ray Research, University of Tokyo, Japan Society for Promotion of Science, and the Kashima Foundation, and in the USA by NASA (Space Physics), NSF (Particle Physics, Polar Programs, EPSCoR), DOE (High Energy), and by the State of Alabama EPSCoR program; and by the Polish State Committee for Scientific Research.

| | **Exposure factor** | **Number of events above $\Sigma E_\gamma \geq 20$ TeV** | | |
|---|---|---|---|---|
| | | C~O | Ne~S | Z ≥ 17 |
| **JACEE-1~8** | 289 m$^2$hr | 8 | 5 | 6 |
| **JACEE-10** | 49 m$^2$hr | 0 | 2 | 1 |
| **JACEE-11** | 260 m$^2$hr | yet to be recovered | | |
| **JACEE-12** | 212 m$^2$hr | 8 | 5 | 4 |
| **JACEE-13** | 361 m$^2$hr | the analysis is now in progress. | | |

**TABLE 1** *The number of events of each charge group, C–O, Ne– S, and $Z \geq 17$ above the selection threshold energy of $\Sigma E_\gamma \approx 20$ TeV for different exposures. The exposure factor of JACEE-12 corresponds to 5/6 of its total exposure(254 m$^2$hr).*